# Main Manuscript for

One-dimensional van der Waals Heterostructures: Growth Mechanism and Handedness Correlation Revealed by Non-destructive TEM

Yongjia Zheng[a,1], Akihito Kumamoto[b,c,1], Kaoru Hisama[a], Keigo Otsuka[a], Grace Wickerson[d,2], Yuta Sato[e], Ming Liu[a], Taiki Inoue[f], Shohei Chiashi[a], Dai-Ming Tang[g], Qiang Zhang[h], Anton Anisimov[i], Esko I. Kauppinen[h], Yan Li[j], Kazu Suenaga[e,k], Yuichi Ikuhara[c], Shigeo Maruyama[a,3], Rong Xiang[a,3]

1 Y.Z. and A.K. contributed equally to this work.

2 Present address: Department of Materials Science and Engineering, Northwestern University, Evanston, IL 60208

3 Shigeo Maruyama, Rong Xiang.

**Email:** maruyama@photon.t.u-tokyo.ac.jp or xiangrong@photon.t.u-tokyo.ac.jp

**Author Contributions:** R.X. and S.M. designed research; Y.Z., A.K., K.H., G.W., M.L., Q.Z., A.A., R.X. performed research; Y.Z. synthesized 1D heterostructures; R.X. and A.K. took HR-TEM images and ED patterns; A.K. took STEM images and EELS mappings; R.X. analyzed edge structures; A.K. and R.X. performed TEM image simulations; T.I. synthesized horizontally aligned SWCNT arrays; G.W. performed cross-sectional TEM; T.I. fabricated TEM grids; R.X. Y.Z. and M.L. performed EDS mappings and analyzed nucleation sites; Q.Z., A.A. and E.K. synthesized large-area SWCNT films; K.H. performed MD simulations; K.O. S.C. and S.M. analyzed simulation results; Y.S. and K.S. advised TEM experiments of handedness relationship; D.T. advised TEM image interpretation; R.X. built atomic models and made illustrations; R.X. and Y.Z. wrote the paper; all co-authors joined the discussion and approved the submission of the manuscript.

**Competing Interest Statement:** The authors declare no competing interests.

**Classification:** Physical Sciences, Applied Physical Sciences

**Keywords:** one-dimensional van der Waals heterostructure, carbon nanotube, transmission electron microscopy.

**This PDF file includes:**

    Main Text
    Figures 1 to 6
    Tables 1 to 2

**Abstract**

We recently synthesized one-dimensional (1D) van der Waals heterostructures, in which different atomic layers (e.g. boron nitride, or molybdenum disulfide) seamlessly wrap around a single-walled carbon nanotube (SWCNT) and form a coaxial, crystalized heteronanotube. The growth process of 1D heterostructure is un-conventional – different crystals need to nucleate on a highly curved surface and extend nanotubes shell by shell – so understanding the formation mechanism is of fundamental research interest. In this work, we perform a follow-up and comprehensive study on the structural details and formation mechanism of chemical vapor deposition (CVD) synthesized 1D heterostructures. Edge structures, nucleation sites, crystal epitaxial relationships are clearly revealed using transmission electron microscopy (TEM). This is achieved by the direct synthesis of heteronanotubes on a CVD-compatible Si/SiO$_2$ TEM grid, which enabled a transfer-free and non-destructive access to many intrinsic structural details. In particular, we have distinguished different shaped boron nitride nanotube (BNNT) edges, which are confirmed, by electron diffraction at the same location, to be strictly associated with its own chiral angle and polarity. We also demonstrate the importance of surface cleanness and isolation for the formation of perfect 1D heterostructures. Furthermore, we elucidate the handedness correlation between SWCNT template and BNNT crystals. This work not only provides an in-depth understanding of this new 1D heterostructure material group, but also, in a more general perspective, serves as an interesting investigation on crystal growth on highly curved (radius of a couple of nm) atomic substrates.

**Significance Statement**

We recently synthesized coaxially nested one-dimensional van der Waals heterostructures, in which boron nitride nanotubes or molybdenum disulfide nanotubes grew seamlessly on a single-walled carbon nanotube template. In this work, edge structures, nucleation sites, crystal epitaxial relationships in heteronanotubes are unambiguously revealed by a non-destructive transmission electron microscopic technique. These understandings, together with the characterization technique developed here, can help to optimize the synthesis process. Structure-controlled heteronanotubes may, ultimately, be used to build nanoscale devices such as gate-all-around nanotube transistors.

**Introduction**

Two dimensional (2D) atomic layer materials including graphene, hexagonal boron nitride (BN), transition metal dichalcogenide (TMDC) have emerged as a research focus in the past decades.(1-4) They represent a new class of crystals that were believed to be unstable before, and therefore their discoveries allowed exploration of new physics and development of novel devices. Van der Waals (vdW) heterostructures, which are artificial stacks of these different atomic layers, further extended the research scope of 2D materials.(5) These heterostructures offer a nearly unlimited freedom to combine crystals beyond conventional limits of crystal type and lattice matching.(6) This generated a lot of interesting progress from fundamental crystallography, physics, to optical and electronic devices.(7-11) However, the concept of the vdW heterostructures has mostly been investigated in 2D materials.

Earlier, we demonstrated the experimental synthesis of one-dimensional (1D) van der Waals heterostructures.(12) BN nanotubes (BNNT), molybdenum disulfide nanotubes (MoS$_2$NT) were synthesized by chemical vapor deposition (CVD) on a single-walled carbon nanotube (SWCNT) template. As a result, coaxial heterostructures consisting of different nanotube crystals were generated. The growth of these heteronanotubes were shell by shell, a very different growth process from conventional growth of 1D homo-material nanotubes where multiple walls are formed simultaneously from a nanoparticle. Studying nucleation and crystal growth behaviors in 1D vdW heterostructures is of fundamental research interest but is challenging as all processes occur on tiny (only a couple of nm) and highly curved surfaces.(13, 14)

Transmission electron microscopy (TEM) is one of the most powerful tools to study the atomic arrangement of a nanomaterial. Recent advances even enabled the identification of local electronic state, or even phonon state in a crystal.(15, 16) However, one well-recognized obstacle is that TEM observation usually requires a complex sample preparation process. This preparation process not only makes TEM characterizations inefficient, but also often results in a change of the material geometry or loss of intrinsic sample information. Recently we developed an approach using a high-temperature-stable $Si/SiO_2$ TEM grid.(17) This grid is fabricated by MEMS techniques starting from a $Si/SiO_2$ wafer. It can be used to support high temperature reactions in a way the same as conventional Si wafers but, at the same time, is also compatible to advanced TEM observations. This routine allowed us to observe the geometry and the atomic arrangement of a nanomaterial efficiently without sample processing. Most importantly, it becomes possible to keep and reveal the original geometry, structure details and even the evolution after high temperature processing.

In this study, we combine this TEM approach with the growth of 1D vdW heterostructures and investigate the growth processes of BNNTs on SWCNT templates. Benefitting from the transfer-free strategy, the heterostructures were kept in their most intrinsic morphology, and many structural details can be clearly visualized. Several fundamental issues such as edge structure, nucleation behavior, requirements for a perfect growth are then able to be unambiguously illustrated. Furthermore, we investigated the handedness correlation between grown nanotube(s) and the template, which is a unique geometric feature for 1D multilayers, yet was very difficult to investigate and thus barely understood. Global and local energy difference between uni- and contra-handed heterostructures were calculated to present some theoretical insights into the experimentally observed semi-epitaxial relationship.

## Results and discussion

### 1. Overview of our TEM approach and 1D vdW heterostructure

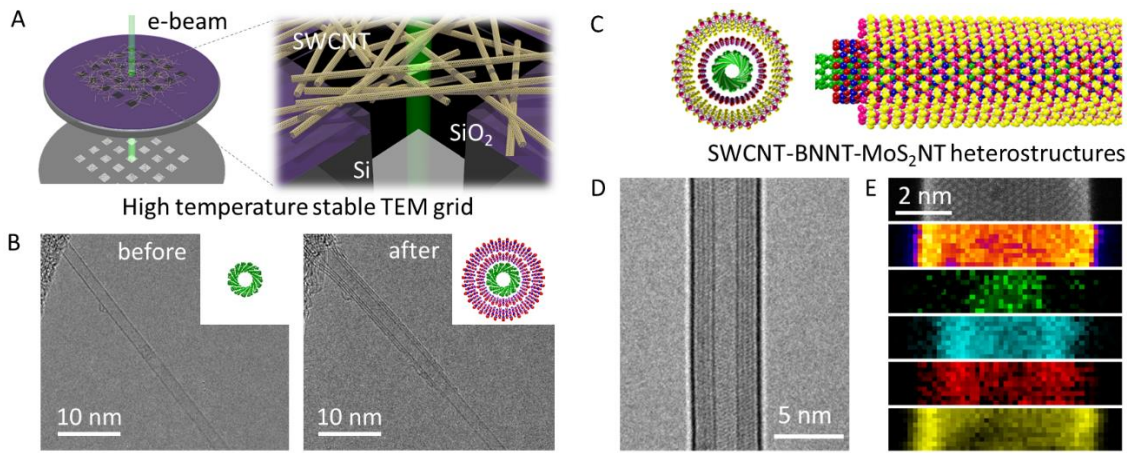

**Fig. 1.** The experimental strategy showing the growth of 1D vdW heterostructures directly on TEM grid. (A) Schematic of the high-temperature-stable TEM grid. (B) TEM images of the exact same SWCNT before and after the growth of outer BNNT. (C) Atomic model, (D) TEM image and (E) EELS mapping of a SWCNT-BNNT-$MoS_2$NT heterostructure (green, blue, red, yellow indicate the elemental distribution of carbon, boron, nitride, sulfur, respectively).

**Fig. 1** shows an overview of the TEM strategy we used, as well as the structure of the 1D vdW heterostructures that we synthesized directly on our high-temperature-stable TEM grids. The TEM grid is made of Si and $SiO_2$, with a thin $SiO_2$ layer as support. In this study suspended $SiO_2$ is also etched so that there are empty windows that can support SWCNT networks (Fig. 1A). Because this grid is stable up to 1100°C, we can use it directly in a CVD growth furnace. The grid is compatible with use in most TEM holders. At a slightly reduced beam density, various structural and elemental characterization can be performed directly on this grid without any treatment. This strategy significantly improves the efficiency of TEM to reveal the original geometry and the structure details of our heteronanotubes. With the assistance of coordinates or markers on the grid, we are able to observe the same location (*SI Appendix,* Fig. S1), and even the exact same nanotube before and after a high temperature CVD. One example is shown in Fig. 1B. The original nanotube is single-walled, but after a BN CVD, two to three additional layers appeared on the surface of this SWCNT. By this strategy, the growth of additional BNNTs can be straightforwardly illustrated. When putting a SWCNT-BNNT structure into another CVD process, a ternary heterostructure of SWCNT-BNNT-$MoS_2$NT can also be fabricated. The example in Fig. 1C-E is a coaxial nanotube containing one layer of carbon, three layers of BN and one layer of $MoS_2$. Different materials can be clearly visualized by the image and electron energy loss spectroscopy (EELS) mapping.

## 2. Open edge growth of 1D vdW heterostructures

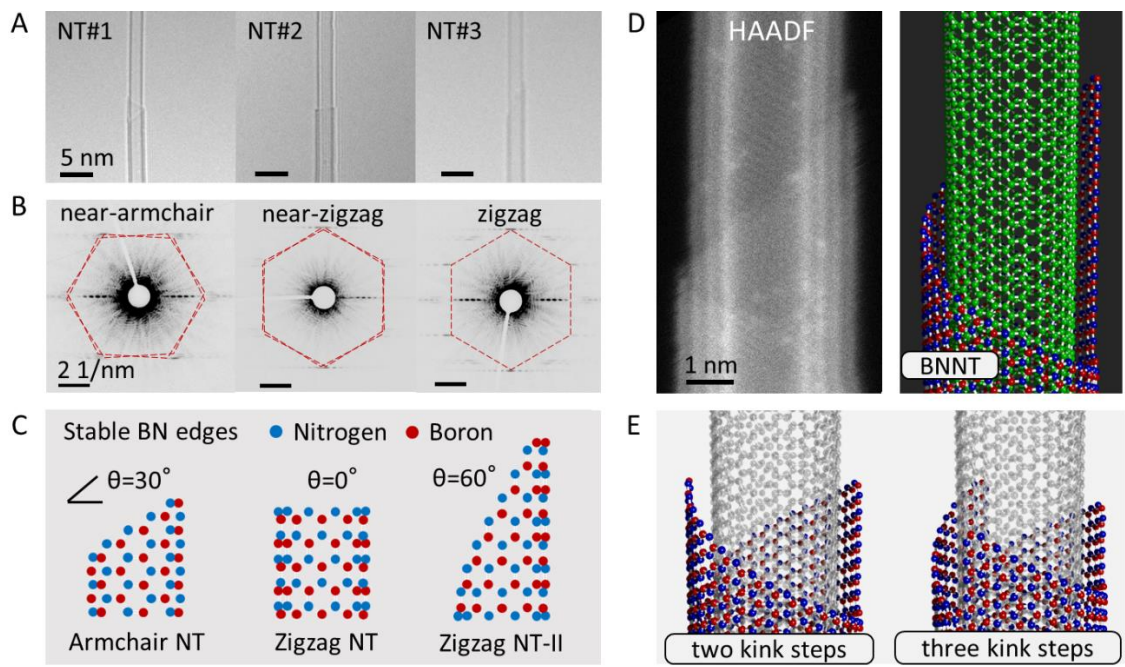

**Fig. 2.** Characterization of the open growing edge of the 1D vdW heterostructure. (A) TEM images of three different heteronanotubes showing each outer BNNT usually has a sharp-cut open edge, but the edge is aligned to the nanotube circumference with different angles. (B) NAED patterns of these corresponding three nanotubes, suggesting the chiral angle of outer BNNTs are 27°, 3° and 0°, respectively. (C) Expected atomic arrangements at the edge of these three BNNTs revealed by NAED patterns, indicating that the sharp-cut edges are N-terminated zigzag BN side. (D) Experimental HADDF-STEM image of a BN edge and its schematic (not exact) atomic structure.

(E) Schematics showing open edges with two or three kink steps. In this case the inner nanotubes are made semi-transparent to show all kinks.

As heterostructures were directly synthesized on Si/SiO$_2$ TEM grid, this "heterostructure-on-grid" sample can be brought from our CVD chamber to the TEM column without further processing. Many intrinsic details of the heterostructure can be thereby preserved. The first thing we noticed is that the open ends of BNNTs can be visualized by TEM. Under-focused images, even by conventional high resolution TEM (HRTEM), can reveal the shape of a tube edge, and this shape is found to be different from tube to tube. Fig. 2A represents three typical types of open edges we observed. These edges of BNNTs usually have sharp-cut shapes but are cut diagonally with various inclination angles from the circumference of the nanotube. For example, NT#1 has a clear spiral end and the edge inclination angle (between the edge and tube circumference) is roughly 30°, while NT#2 has an edge nearly perpendicular to the SWCNT axis. To distinguish the difference between these two tubes, nano-area electron diffraction (NAED) was employed to determine the chiral angle (crystal orientation) of outer BNNTs.(18) NAED patterns in Fig. 2B suggests that NT#1 is a near-armchair BNNT with the chiral angle of 28°, while NT#2 is a near-zigzag BNNT with chiral angle of 2°. Plotting the atomic arrangement of these nanotubes in Fig. 2C immediately reveals that, in both cases, the sharp cuttings we observe correspond to the zigzag edge of a BN honeycomb lattice. That is, an open-ended BNNT in this study prefers a zigzag edge over an armchair edge. This is consistent with previous theoretical and experimental studies in 2D BN where zigzag edge is found to be energetically more favorable.(19-21) With this ability of observing the edge shape and identifying the chiral angle for the same nanotube, we learned that the open end of a BNNT is aligned to its axis with an angle depending on its own chiral index. An armchair BNNT tends to have a spiral edge (NT#1) while a zigzag BNNT has a perpendicular edge (NT#2), both of which are due to the preference of a zigzag edge. This correlation is straightforward and clearly demonstrated by the above characterizations.

However, one must take into account the difference in structural polarity when describing the edge of a BNNT (unlike in CNT where all atoms are carbon), and this polarity explains a second type of edge shape for zigzag BNNTs shown in NT#3. Because of the crystal symmetry, a zigzag BN edge can be either terminated with B (B-polar) atoms or N atoms (N-polar), and studies in 2D material indicated that the N-terminated zigzag edge formation is energetically more stable. Therefore, when a zigzag BNNT is N-polar in its growing direction, a flat cut edge is stable as shown in NT#2. However, when a zigzag BNNT is B-polar in its growing direction, a flat-cut edge becomes unstable and must grow to a 60° inclination angle to form the N-termination. This is exactly the case for NT#3. The chiral index of this BNNT is zigzag (34, 0), as suggested by the NAED pattern. Its chiral angle, 0°, is nearly the same as NT#2 but the edge shape is completely different. This indicates that the polarity of open ends in NT#2 and NT#3 are opposite, with NT#2 N-polar up and NT#3 B-polar up. Therefore, more generally, chirality of the BNNT, together with the polarity of the tube-growing direction, determines the stable edge structure of outer BNNTs. More discussions on this point, including the comparison between graphene and hexagonal BN (*SI Appendix,* Fig. S2),(22) edge structure of a chiral BNNT (*SI Appendix,* Fig. S3) and the method to extract the diffraction of BNNT from the two sets of patterns (*SI Appendix,* Fig. S4), are provided in the supporting information.

Scanning TEM (STEM) was actively employed in the past decade to perform atomic-scale analysis for material science. Especially, high-angle annular dark-field (HAADF) STEM imaging is a critical tool to evidence foreign heavy atom inclusion inside materials. One representative HAADF STEM image of an open end of a BNNT is shown in Fig. 2D. First, this HAADF images further supports that the tube end has a zigzag edge, as a lattice space contrast of 0.21 nm appears in this image and is about 30° to the tube circumference. A second point we can elucidate from this HAADF-STEM imaging is that the growth outer BNNT is a metal free process. Although there are always small amounts of Fe remaining in the starting SWCNT network, (23) it seems that no Fe atoms served as the catalyst for the extension of BNNT in the current growth scenario. This can be evidenced by STEM image simulation. If we add one Fe to any atomic position of our structure model, a very bright point contrast will appear. By comparing experimental and simulated images we can conclude that the heteronanotubes followed a metal free growth model. This is also supported by the very slow growth of the outer BNNTs. The growth rate is about several nanometers per minute, which is three to five orders of magnitude slower than catalytic growth of 1D or 2D BN.(24, 25) In addition, in all our observations, every BN edge has only one major kink step, rather than multiple zigzag kinks steps (Fig. 2E). It is still unclear at this stage, but number of kink steps were expected to affect the growth dynamics of the open-growing nanotube, as has been predicted by a previous theory in carbon nanotube.(14) It may be interesting to study the growth dynamics for different edges in the future.

**3. Essential requirements for the formation of a perfect heterostructure.**

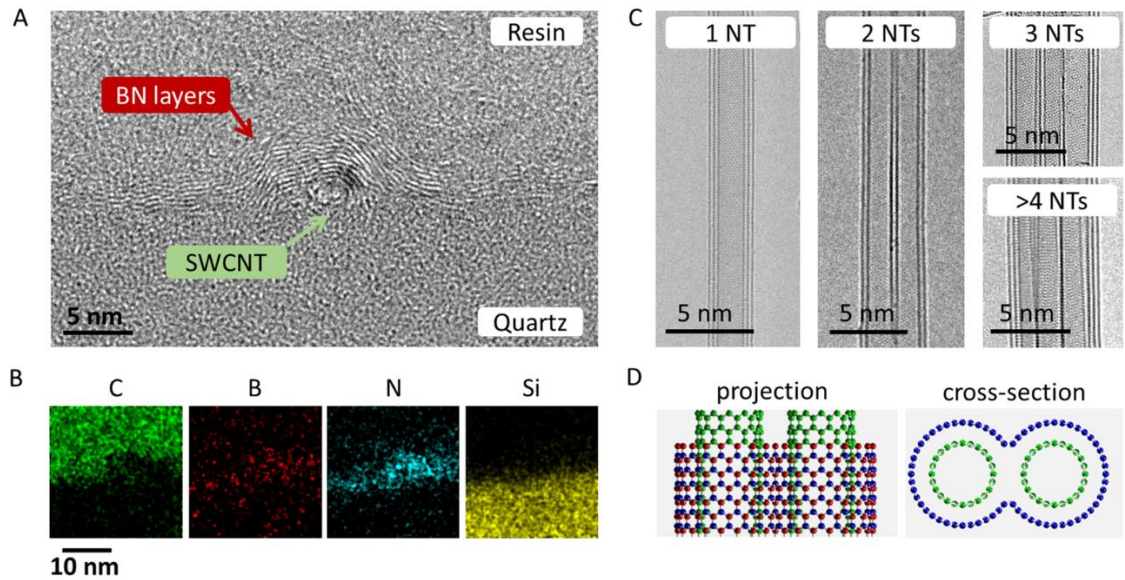

**Fig. 3.** Importance of isolation. (A) Cross-sectional TEM image and (B) EDS elemental mapping of BN coating on a SWCNT sitting on a quartz substrate, suggesting BN covers only top half of the SWCNT and cannot detach SWCNT from the substrate to form a seamless cylinder. (C) TEM images of the one to two layer BN coating on an individual SWCNT, a two-NT bundle, a three-NT bundle and a larger bundle, confirming BN wrap the entire bundle other than isolating the nanotubes inside a bundle. (D) Expected projection and cross-sectional view of the BN wrapping on a bundle.

The growth of outer BNNTs started from the nucleation on the surface of an SWCNT template, and we confirm that isolation and surface cleanness are two essential requirements for the successful formation of perfect heterostructures. Requirement for an isolation is straightforward to understand and Fig. 3 shows two situations of BN coating when nanotubes are not well isolated. In the first case as shown in Fig. 3A and B, a single SWCNT is sitting on a quartz substrate, and cross-sectional TEM reveals that the BN coating covers the upper half of the SWCNT, rather than lifting the SWCNT to generate a seamless, cylindrical structure. A gentle but distinguishable contrast difference between upper resin and lower quartz can be observed in Fig. 3A, and energy-dispersive X-ray spectroscopy (EDS) in Fig. 3B further verifies the existence of an interface, where a SWCNT and multiple BN layers locate. In another case, if several SWCNTs are touching each other in a bundle form, BN coating covers the outer surface of the entire bundle, and is unable to separate these SWCNTs and form perfect heterostructures on each SWCNTs (Fig. 3C and 3D).

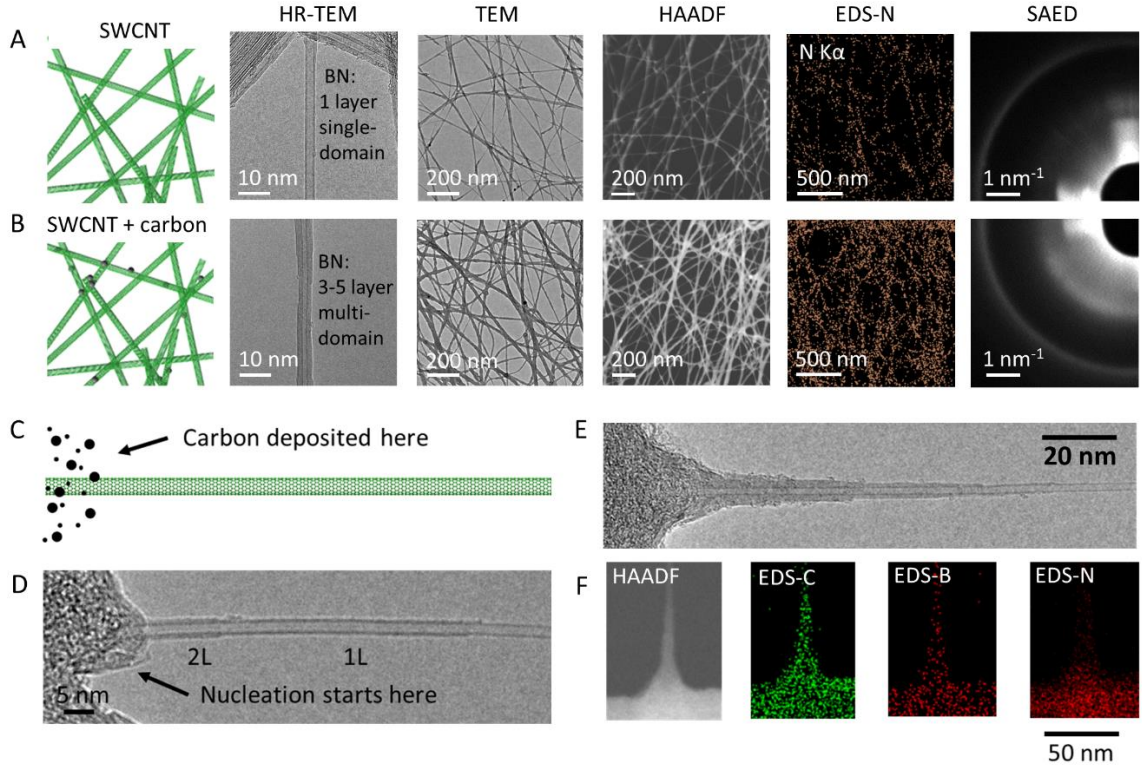

**Fig. 4.** Importance of cleanness. A comparison of BNNT growth behaviors on (A) clean SWCNT networks and (B) SWCNT networks with intentionally deposited amorphous carbon, suggesting that BNNT can nucleate from single site and grow into a perfect crystal on clean SWCNT surface, while multiple nucleation and polycrystalline growth occurs when there are contaminations on SWCNT surface. (C-F) This strategy may be used to achieve a site-selective growth, as revealed by TEM imaging and EDS mapping.

The second essential requirement for the formation of a heteronanotube is the surface of the template SWCNT. Perfect heterostructures can only be formed on an ultra-clean SWCNT. Fig. 4 shows the different coating behaviors for BN on a clean SWCNT network and an amorphous-carbon-contaminated SWCNT network. All the TEM, STEM, EDS, and selected area electron diffraction (SAED) analyses were performed under the exactly same machine condition to achieve a strict comparison. On clean SWCNT surface, usually one or two layers of BN are formed from a single starting point, while on carbon-contaminated SWCNT surfaces, BN usually nucleates from multiple sites and grows into a multi-layer, poly crystal nanotube. Over the whole area, the BN signals are much stronger on carbon-contaminated SWCNT networks than the clean ones, suggesting that intentionally-deposited carbon or other unintentional contaminations may serve as nucleation sites of the outer BNNT growth. Therefore, in order to obtain perfect heteronanotubes, it is indispensable to avoid contaminations to SWCNT templates. In our process this is achieved by using an ultraclean SWCNT. Furthermore, the use of our low pressure CVD chamber also contributes to the formation of high-quality 1D heterostructures.

From another perspective, this finding can be used to achieve a site-selective growth of heterostructures, which may be an important step for electronic device or optical characterizations. One example is demonstrated as Fig. 4C-F. In this experiment, we intentionally deposited some carbon using e-beam at a small local area of a clean SWCNT before putting this nanotube into the BN CVD. As expected the growth started from the position where carbon is deposited, as can be evidenced by TEM, STEM-HAADF imaging, and EDS mapping.

Without intentional deposition, BNNT growth usually starts from the tube-tube contacts (shown in *SI Appendix,* Fig. S5), where BN precursors may accumulate more easily. This phenomenon is particularly clear when an individual-tube-enriched SWCNT network is used as the starting growth template.(26) Nucleation of BNNT from tube-tube contacts usually results in different modes, one-end nucleation or both-end nucleation on an isolated nanotube. In a both-end nucleation, two outer BNNTs can have different chiral indexes, although they are formed on the same SWCNT. They can even meet each other to form a junction when the CVD time is long enough. This is conclusively evidenced by NAED patterns, or by the different wall contrast (to be further described in the next section). Occasionally we also observe the growth of BNNT at the center of an isolated nanotube, but this center-nucleation mode is less predominant. The summary of these three different nucleation modes, and an aberration corrected image of a BNNT junction are provided in the supporting information (*SI Appendix,* Fig. S6).

## 4. Handedness relation in 1D vdW heterostructures

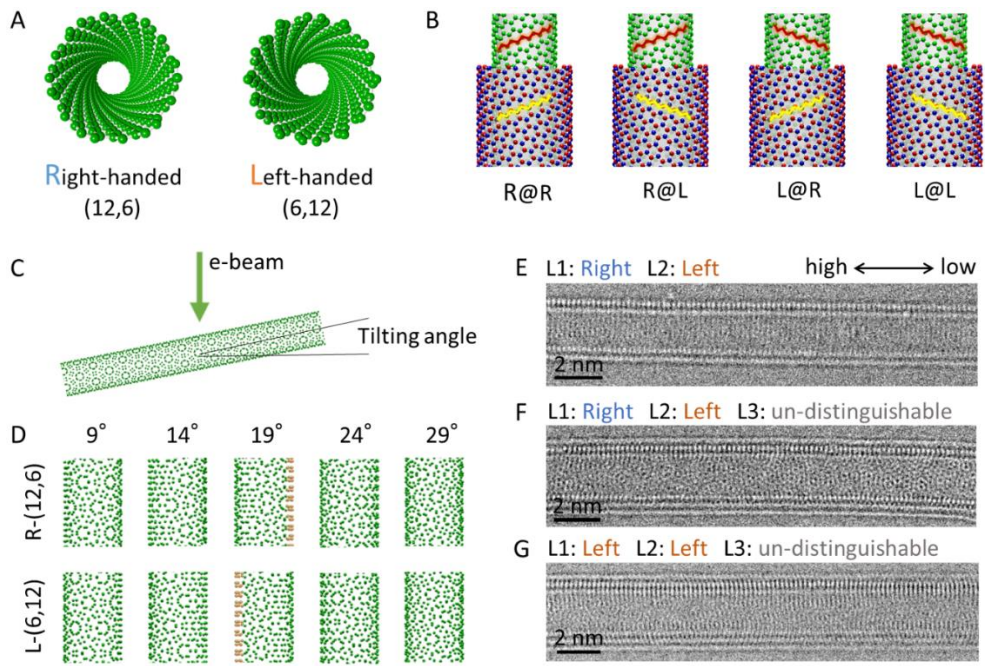

**Fig. 5.** Handedness relationship in 1D SWCNT-BNNT vdW heterostructures. (A) Geometry and definition of right-handed and left-handed nanotube. (B) In a double-walled SWCNT-BNNT heterostructure, four different handedness combinations, R@R, R@L, L@R, and L@L, can appear. (C,D) The experimental approach to identify the handedness of a SWCNT or BNNT nanotube in a TEM. When a nanotube is tilted, dot-like contrast appears at one side of the wall depending on the handedness of the nanotube. (E-G) Representative TEM images of three SWCNT-BNNT heterostructures having different handedness combinations.

Previously we have confirmed that there is no obvious chirality correlation between the inner SWCNT and outer BNNT in a heterostructure,(12) but the handedness correlations still remain unknown. Handedness is a unique structural feature for 1D tubular nanostructures and the experimental identification is extremely difficult.(27, 28) It is usually recognized as the ultimate structure characterization for a nanotube, as a left-handed nanotube and a right-handed nanotube have a mirror symmetry for their atomic arrangements, but are supposed to have exactly the same electronic and chemical properties (Fig. 5A).(29) There are only a couple of techniques that have been proven capable of identifying handedness.(30-33)

In TEM imaging, a right-handed (12,6) nanotube and left-handed (6,12) nanotube show similar contrasts. Electron diffractions, although capable of telling the chiral index of a nanotube, cannot distinguish a left-handed nanotube from right-handed one as their patterns are exactly the same. Despite the technical challenges, identifying the handedness is scientifically meaningful particularly in the case of a heterostructure. One reason is that, if one would unfold a double-walled heteronanotube (should have four different combinations as shown in Fig. 5B), even for the same configuration, e.g. (17,13)@(34,0), a left-left handed (L@L) and left-right handed (L@R) have completely different stacking angles and thus Moiré patterns. Considering the recent progress in moiré physics (e.g. in both graphene-graphene and graphene-hBN) for 2D crystals,(34) and more recently 1D double-walled carbon nanotube (e.g. moiré-induced tube-tube coupling and new states),(35) it is not only important to satisfy fundamental scientific curiosity in crystallography, but also of great importance to understand how the properties of a heteronanotube may be affected by their atomic arrangements.

A simple TEM technique for the identification of handedness in 1D nanotubes is to use side-wall contrast.(36) In brief, when a nanotube is tilted to a certain angle in TEM, a discontinuous, dots-like contrast, which originates from 0.21 nm zigzag atomic chain of the hexagonal honeycomb lattice, will appear at one side of projected TEM image depending on which handedness the nanotube is. Fig. 5C and 5D shows the scheme of this method. Historically when this method was proposed, a nanotube needed to be tilted to its nearly exact chiral angle to see a clear side contrast. For example, a (12, 6) nanotube needs to be tilted to 19° in order to see this side-wall armchair contrast. However, with recent development of aberration-corrected TEM and hence the improvement of the point resolution, we notice that even within ±5°, side-wall contrast of a nanotube can be observed in our aberration-corrected TEM. For a (12,6) nanotube having a chiral angle of 19°, within the range when the tube is tilted from 14 to 24°, the sidewall contrast can be distinguished. This decently large angle tolerance suggests that, in a SWCNT network where chiral indexes of SWCNTs are randomly distributed (as confirmed previously), we can roughly identify the handedness of 1/3 of the nanotubes among the whole population by only tilting the TEM holder to one angle. This is a convenient way to identify many nanotubes in the sample or many walls in one multi-walled heteronanotube. Occasionally we can also observe 0.12 nm armchair contrast at the wall (but this requires the nanotube to be tilted in a very narrow angle range), which can help to further identify more nanotubes. Detailed explanation to this method, including the mechanism and examples of the side wall contrast are provided in supporting information (*SI Appendix,* Fig. S7).

Representative experimental data are shown in Fig. 5E-G. In all these three images, the sample is tilted as the left side is higher in TEM and the right side is lower. In the first image, clear contrast appeared at the bottom wall for the inner tube and at the upper wall for the outer BNNT, which suggests that the inner nanotube is a right-handed and the outer nanotube is a left-handed.

The second example in Fig. 5F is a triple-walled heteronanotube. The inner and middle shells also have a contra-handedness, but the outer shell is undistinguishable as no obvious contrast appears at either side of the walls. This indicates the chiral angle of this outmost shell is beyond the range of tilting angle ±5°. The third heteronanotube however has a uni-handedness. Both the inner and middle shells are left-hand, while the outer shell is also undistinguishable. We listed the number 1 to 10 nanotubes we studied as Table 1 and the whole data set can be found in supporting information (*SI Appendix,* Table S1).

**Table 1. Summary of the handedness information in all nanotube shells**

| No. of NT | 1 | 2 | 3 | 4 | 5 | 6 | 7 | 8 | 9 | 10 |
|---|---|---|---|---|---|---|---|---|---|---|
| 1st layer (C) | R* | R | L | R | Z | Z | Z | Z | R | Z |
| 2nd layer (BN) | L | L | L | R | L | L | R | Z | R | L |
| 3rd layer (BN) |   | ? | ? | R | L |   | R | Z | ? | L |
| 4th layer (BN) |   |   |   | R | L |   |   | L |   |   |
| Summary | Left-handed | | Right-handed | | Near-zigzag | | Near-armchair | | | |
| Counts | 52 | | 44 | | 58 | | 2 (TEM-limited) | | | |

*R: right-handed    L: left-handed    Z: near-zigzag

As a brief conclusion for all the nanotubes (meaning all distinguishable shells in all heteronanotubes we studied), we found 52 left-handers and 44 right-handers. It is reasonable that two types of nanotubes are half and half. In addition, there are also a decently large number of near-zigzag nanotubes observed. However, there are only two near-armchair nanotubes picked up by TEM. This is a limitation of TEM and does not reflect the real abundance of near-armchair nanotubes inside the whole population.

When comparing the handedness correlation for different shells within a same heteronanotube, a summary is presented in Table 2. First, between the first shell (SWCNT) and the second shell (BNNT), we found 27 sets having a uni-handedness, while 22 sets having a contra-handedness. The numbers are enough to prove that there is no obvious correlations between the SWCNT and grown BNNT. However, when counting the second shells and beyond, a weak tendency of having the same handedness begin to appear. To be specific, among the 51 sets of data we obtained, 37 pairs are uni-handed.

**Table 2. Summary of the correlation between inner and outer nanotubes**

| 1st interlayer (SWCNT-BNNT) | | 2nd interlayer and beyond (BNNT-BNNT) | |
|---|---|---|---|
| same | different | same | different |
| 27 | 22 | 37 | 14 |
| 55.1% | 44.9% | 72.5% | 27.5% |

To understand the mechanism behind this TEM obtained results, we performed a theoretical calculation with empirical potentials on the different combinations of SWCNT and BNNTs. We started from a uni-handed and a contra-handed double-walled BNNT. Fig. 6A illustrates the atomic energy distribution of these two structures. Details of the calculated energy values are available at *SI Appendix,* Table S2 and S3. The first thing we learned is that the global energy difference between a uni-handed and a contra-handed heterostructure is negligible. Although atomic configuration and therefore local energy are apparently different place by place (as revealed in Fig. 6B and C), the whole nanotube is, to our surprise, energetically averaged, causing the absence of any structural preference (unfolded view in *SI Appendix,* Fig. S8). This indicates that, the experimentally observed tendency of having a uni-handedness in the second layer and beyond, are not because the uni-handedness are energetically more stable in equilibrium. However, the growth of heteronanotubes is a dynamic process, and it usually started from nucleating a small site on the surface of existing nanotube. Therefore, these locally formed small sites/fakes, if having any energy preference, will determine the chirality and handedness of the resulting material before a seamless circumference is formed.

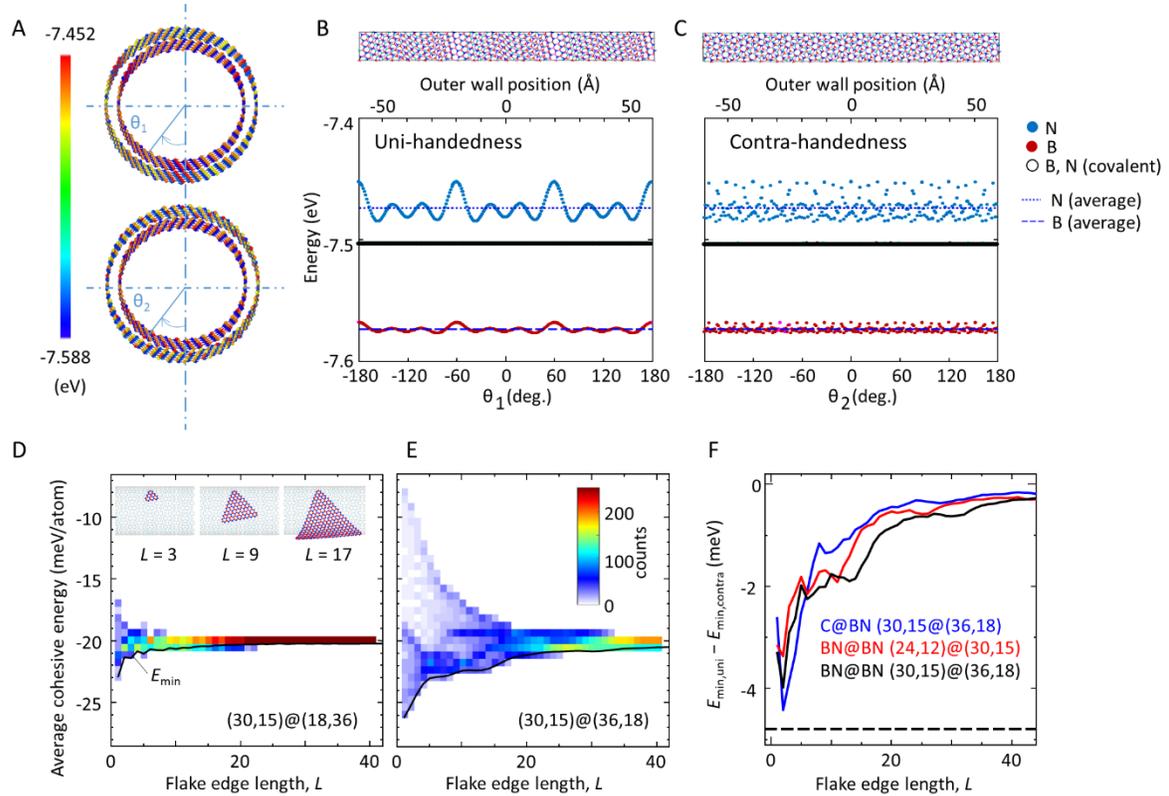

**Fig. 6.** Simulated global and local energy difference in uni- and contra-handed heterostructures. (A) A color map of interatomic potentials of each atom, where $\theta_1$ denotes the circumferential position for double-walled (DW)-BNNT (30,15)@(36,18) and $\theta_2$ denotes the position for (30,15)@(18,36). (B) Interatomic potential of each atom as a function of the circumferential position $\theta_1$. (C) Interatomic potential of each atom as a function of the circumferential position $\theta_2$. Magenta, cyan and black points in (B) and (C) denote potential energy of N, B and both with only covalent bonds, respectively. Blue dashed and dotted lines denote averaged potential energy per atom for B and N atoms, respectively. Histograms for the cohesive energy of a triangular BN flake with various edge lengths $L$ in a (30,15)@(36,18) (D) and a (30,15)@(18,36) (E) BNNT. In total, 252 positions are considered as a flake vertex, which correspond to all B atoms in a unit cell. The energy of each flake is averaged for all B and N atoms that compose the flake. Black lines represent the minimum cohesive energy $E_{min}$ among the 252 positions considered. (F) Difference in $E_{min}$ versus $L$ for uni- and contra-handedness. Horizontal dashed line represents the difference of the cohesive energy in the case of AA' bilayer BN and 19.1° twisted bilayer BN. When the diameter of template NTs is infinitely large, the solid lines in F should be close to the dashed line until some flake size.

To deliver this concept of "local preference", we plotted the average cohesive energy of a triangle flake versus its flake length ($L$ = number of hexagons at each side of the triangular flake) for contra-handedness and uni-handedness in Fig. 6D and 6E. When the flake is placed randomly on the surface of SWCNT, the energy varies with the position, which causes an energy distribution in Y-axis. By this approach, the local energy difference can be clearly visualized -- the local energy distribution is broader in a uni-handedness (Fig. 6E) than a contra-handedness combination, but exact values at some locations are smaller, meaning these locations are energetically more favorable. Small flakes may prefer to nucleate at these "comfortable locations" to form uni-handed seeds which finally further grow into uni-handed nanotube. In short, although there are more "un-comfortable" locations in a uni-handed case, there are also more "very comfortable locations" for nucleation, particularly at the early stage when the flake is small.

This local preference is more significant beyond second layer probably because of two possible factors. The first factor is the material effect. Materials starting from layer two are homogenously BN, and therefore may prefer uni-handedness or even similar chiral angles. This material effect could be smaller between SWCNT and BNNT. The second, and more likely factor is the diameter effect. When one additional layer forms on a nanotube, the diameter increases about 0.7 nm. For larger diameter nanotubes, outer surface is less curved. Local preference on less curved surface will be more like to be more significant, behaving more similarly to local epitaxial nucleation to a flat 2D surface.

We performed the same calculations on a double-walled CNT-BNNT heterostructure, and a smaller diameter BNNT-BNNT homostructure, and extracted their conformable energy spots ($E_{min}$ value). The difference between uni-handness and contra-handedness ($E_{min,uni}$ - $E_{min,contra}$) is plotted in Fig. 6F and compared with the previous double-walled BNNT-BNNT. Although it is still hard to quantify the material effect and curvature effect for a large number of different pairs, and in a more collective way, material effects are clearly observed (blue line vs. black line in Fig. 6F). Therefore, the combinational effect of these two factors is expected to determine the extent of inter-wall epitaxy and explains the tendency of handedness correlation observed in our TEM characterizations. However, diameter effect may be more determinative. Ultimately when the curvature disappears, 2D BN flakes are usually grown epitaxially following the orientation of a graphene substrate.(24) Despite that further efforts are still needed to fully understand this tendency, the random correlation on the first shell, and the later preference of uni-handedness on second layer and beyond, seems to be convincing and conclusive in the study. This may help us to understand the shell-shell interaction, and thereby how the optical, electronic, topological and other properties of a heteronanotube may be modulated.(37-42) Possibly it will also be beneficial for the design of different and new heterostructures for devices.(13, 43, 44)

**Conclusion**

In summary, we have performed a systematic study on the growth mechanism and the structure correlation in our newly developed 1D vdW structures. Benefited from the CVD-compatible TEM approach, the original geometry and intrinsic structural details of the heterostructures are preserved and become conveniently accessible. The key findings are summarized as follows. (i) 1D SWCNT-BNNT vdW heterostructures follows an open-end growth model, and the end of an outer BNNT has a sharp-cut edge which is aligned to the tube circumference at an angle depending on its own chiral index and polarity. The mechanism behind this phenomenon is that BNNT open edges always prefer N-zigzag termination, as were evidenced by NAED patterns and direct TEM/STEM imaging. (ii) Having a well-isolated, ultra-clean SWCNT is essential to obtain a perfect heterostructure, as BN will cover the only half of the tube or over the entire bundle if the tube is attaching to a substrate or other nanotubes. Furthermore, cleanness of the surface is also essential to start a single site nucleation and form a single crystal domain on the nanotube. However, intentionally introduced nucleation sites may be utilized to achieve a site-selective fabrication of heterostructures. The nanotubes usually nucleate from the tube-tube contacts, and this resulted in three different growth cases: one-end growth, two-end growth and center growth. (iii) The cutting-edge aberration-corrected TEM allowed to identify the handedness of roughly 1/3 of the nanotubes among the whole population. Handedness of the inner and outer nanotubes was confirmed to be random, but there was a gentle tendency toward the same handedness beyond the third layer. This was not because of an energetic preference between different handedness combinations as the global energy difference between uni- and contra-handedness is very small, but probably caused by the local structure preference which happened at the nucleation stage on the surface of the templated nanotube. These effects are expected to be more significant for homo-materials and larger diameter nanotubes.

These comprehensive results were obtained on $Si/SiO_2$ based TEM grids, which can serve as a very effective routine from CVD to TEM to study the atomic structure of an as-grown nanomaterial and to understand its formation behaviors. This TEM approach itself is robust and universal and may be used for other material synthesis or other catalytic processes. The knowledge we learned on the growth mechanism and crystal relationship may be beneficial to understand the properties, especially the tube-tube interaction of our current SWCNT-BNNT-$MoS_2$NT heterostructures, and may be helpful for the controlled synthesis, or even the design of new and more sophisticated 1D vdW heterostructures.

**Materials and Methods**

The template SWCNT film used in this project was synthesized by aerosol CVD.(23) Typically ferrocene was used as the catalyst precursor and CO was used as the carbon source. The growth temperature was 1000-1200°C. The SWCNTs were formed in gas phase and collected onto a filter paper.

The highly individual SWCNTs are prepared in an injection floating catalytic CVD.(26) 8000 standard cubic centimeters per minute (sccm) of $H_2$ carrier gas and 11 sccm of $C_2H_4$ carbon source were introduced, and 4.0 µl min$^{-1}$ of mixed solution [toluene (10 g), ferrocene (0.3 g), and thiophene (0.045 g), acting as a carbon source, catalyst precursor, and growth promoter, respectively] was injected into the reactor by a syringe pump. The growth temperature was set to be 1100°C.

Horizontally aligned SWCNTs were synthesized on crystal quartz substrates by alcohol catalytic CVD.(45, 46) Fe (nominal thickness of 0.2 nm) was prepared by thermal evaporation, and patterned into line shape with 300 µm spacing by e-beam lithography. Typical SWCNT growth were performed at 800°C and with an ethanol pressure of 140-150 Pa.

SWCNT-BNNT heterostructures were synthesized by a low pressure thermal CVD using ammonia borane ($H_3NBH_3$) as the BN precursor.(12) Briefly, the starting SWCNT prepared in the previous section was placed at the center of the furnace. 30 mg BN precursor was loaded at the upstream and heated to 70-90°C. Vapor of BN precursor was taken by a flow of 300 sccm Ar (with 3% $H_2$) to the hot zone to form BNNT on surface of SWCNTs. The reaction temperature was 1000-1100°C and the chamber pressure was maintained at 300 Pa. The coating time in this study varied from a few min to 1 hr.

SWCNT-BNNT-$MoS_2$NT heterostructures were synthesized by a low-pressure CVD using $MoO_3$ and S powder as precursors.(12) S powder were placed at the upper stream and heated to 100-130°C, and $MoO_3$ is placed at next to S and heated to 500-600°C. The vapor was carried in by a flow of 50 sccm Ar to SWCNTs or SWCNT-BNNT heterostructures at the center. The temperature was maintained at 400-600°C, and typical growth time varied from 5-70 min.

Si TEM grid used in this study was prepared by dry etching. Briefly, a Si/$SiO_2$ wafer (thickness of Si 200 µm, thickness of $SiO_2$ 600 nm for both sides) was used as the starting substrate. Photoresist patterns were prepared on the topside of the substrate by photolithography to define window structures and outer shapes of TEM grids. $SiO_2$ and Si were etched from the topside by isotropic reactive ion etching to the depth of ~20 µm. Then, the backside of the substrate was photolithographycally patterned with the similar structure as the topside but with a little larger windows, followed by being etched by deep reactive ion etching. Etching time and cycles were adjusted to completely etch through the Si and SiO2, so that empty windows are formed for TEM observations.

Conventional HRTEM images were taken by a JEM-2800 at an acceleration voltage of 100 kV. EDS and selected area electron diffraction (SAED) patterns of the entire film were taken by the same TEMs with a typical selected area aperture diameter of a few µm and a camera length of 60 cm. Nano area electron diffraction (NAED) patterns of individual SWCNT-BNNT heterostructure were obtained by JEM-ARM200F STEM with a cold field-emission gun operating at 80 kV. In this case a near parallel beam is used together with a small convergence lens aperture (10 µm in diameter) to obtain a small-area electron beam size (~10 nm in diameter). Additionally, the samples were heated at 300 °C during the measurement to avoid carbon contamination using a heating holder (EM-31670SHTH) and a controller unit (EM-08170HCU). HAADF-, ABF-STEM images and corresponding EELS mapping of SWCNT-BNNT-$MoS_2$ heterostructures were obtained in the same TEM or GRAND-ARM™ STEM operating at 80 kV. Aberration corrected TEM images are taken at a different JEM-ARM200F TEM with a cold field-emission gun operating at 120 or 60 kV.

To describe intra-layer covalent bonds, a set of Tersoff potential for B, N, and C materials is applied(47) and to express interlayer interaction, a registry-dependent interlayer potential and screened Coulomb potential is used.(48-50) We used the molecular dynamics package LAMMPS as an implementation of this calculation.(51, 52)

For DW-BNNTs, ($2n$,$n$) BNNT@($2n$+6,$n$+3) BNNT and ($2n$,$n$) BNNT@ ($n$+3,$2n$+6) BNNT are calculated and for CNT@BNNTs, ($2n$,$n$) CNT@($2n$+6,$n$+3) BNNT and ($2n$,$n$) CNT@ ($n$+3,$2n$+6) BNNT are calculated as the example of uni-handedness and contra-handedness, respectively. The structure optimization is carried out under axial pressure at 1 atm for an isolated DW-NT.

**Acknowledgments**

The authors acknowledge Peng-Xiang Hou, Chang Liu, Hui-Ming Cheng for providing the highly isolated SWCNT films, and Hiroyuki Oshikawa for the assistance of TEM sample preparation and observation. Part of this work was supported by JSPS KAKENHI (grant numbers JP18H05329, JP19H02543, JP20H00220, JP20K14660 and JP20KK0114), by JST, CREST grant number JPMJCR20B5, Japan, and by the Academy of Finland project No. 316572. G.W. was supported by Nakatani Research and International Experience for Students Fellowship Program (RIES). Part of the work was conducted at the Advanced Characterization Nanotechnology Platform of the University of Tokyo, supported by the "Nanotechnology Platform" of the MEXT, Japan, grant number JPMXP09A20UT0063 and JPMXP09A21UT0050.

**References**

1. A. K. Geim, K. S. Novoselov, The rise of graphene. *Nat. Mater.* **6**, 183-191 (2007).
2. K. F. Mak, C. Lee, J. Hone, J. Shan, T. F. Heinz, Atomically Thin MoS2: A New Direct-Gap Semiconductor. *Phys. Rev. Lett.* **105**, 136805 (2010).
3. J. D. Zhou *et al.*, A library of atomically thin metal chalcogenides. *Nature* **556**, 355 (2018).
4. K. S. Novoselov *et al.*, Two-dimensional atomic crystals. *P. Natl. Acad. Sci. USA* **102**, 10451-10453 (2005).
5. A. K. Geim, I. V. Grigorieva, Van der Waals heterostructures. *Nature* **499**, 419-425 (2013).
6. A. Koma, Vanderwaals Epitaxy - a New Epitaxial-Growth Method for a Highly Lattice-Mismatched System. *Thin Solid Films* **216**, 72-76 (1992).
7. X. P. Hong *et al.*, Ultrafast charge transfer in atomically thin $MoS_2$/$WS_2$ heterostructures. *Nat. Nanotechnol.* **9**, 682-686 (2014).
8. C. H. Lee *et al.*, Atomically thin p-n junctions with van der Waals heterointerfaces. *Nat. Nanotechnol.* **9**, 676-681 (2014).
9. H. Fang *et al.*, Strong interlayer coupling in van der Waals heterostructures built from single-layer chalcogenides. *P. Natl. Acad. Sci. USA* **111**, 6198-6202 (2014).
10. Y. Liu *et al.*, Van der Waals heterostructures and devices. *Nat. Rev. Mater.* **1**, 16042 (2016).
11. K. S. Novoselov, A. Mishchenko, A. Carvalho, A. H. C. Neto, 2D materials and van der Waals heterostructures. *Science* **353**, aac9439 (2016).
12. R. Xiang *et al.*, One-dimensional van der Waals heterostructures. *Science* **367**, 537-542 (2020).
13. R. Xiang, S. Maruyama, Heteronanotubes: Challenges and Opportunities. *Small Sci.* **1**, 2000039 (2021).

14. F. Ding, A. R. Harutyunyan, B. I. Yakobson, Dislocation theory of chirality-controlled nanotube growth. *P. Natl. Acad. Sci. USA* **106**, 2506-2509 (2009).
15. J. A. Hachtel *et al.*, Identification of site-specific isotopic labels by vibrational spectroscopy in the electron microscope. *Science* **363**, 525-528 (2019).
16. R. Senga *et al.*, Position and momentum mapping of vibrations in graphene nanostructures. *Nature* **573**, 247-250 (2019).
17. H. An *et al.*, Atomic-scale structural identification and evolution of Co-W-C ternary SWCNT catalytic nanoparticles: High-resolution STEM imaging on SiO2. *Sci. Adv.* **5**, eaat9459 (2019).
18. J. M. Zuo *et al.*, Coherent nano-area electron diffraction. *Microsc. Res. Techniq.* **64**, 347-355 (2004).
19. M. Maruyama, S. Okada, Energetics and Electronic Structure of Triangular Hexagonal Boron Nitride Nanoflakes. *Sci. Rep.* **8**, 16657 (2018).
20. Y. Y. Liu, S. Bhowmick, B. I. Yakobson, BN White Graphene with "Colorful" Edges: The Energies and Morphology. *Nano Lett.* **11**, 3113-3116 (2011).
21. Z. H. Zhang, Y. Y. Liu, Y. Yang, B. I. Yakobson, Growth Mechanism and Morphology of Hexagonal Boron Nitride. *Nano Lett.* **16**, 1398-1403 (2016).
22. V. I. Artyukhov, Y. Y. Liu, B. I. Yakobson, Equilibrium at the edge and atomistic mechanisms of graphene growth. *P. Natl. Acad. Sci. USA* **109**, 15136-15140 (2012).
23. A. G. Nasibulin *et al.*, Multifunctional Free-Standing Single-Walled Carbon Nanotube Films. *ACS Nano* **5**, 3214-3221 (2011).
24. H. Arai, T. Inoue, R. Xiang, S. Maruyama, S. Chiashi, Non-catalytic heteroepitaxial growth of aligned, large-sized hexagonal boron nitride single-crystals on graphite. *Nanoscale* **12**, 10399-10406 (2020).
25. R. F. Zhang *et al.*, Growth of Half-Meter Long Carbon Nanotubes Based on Schulz-Flory Distribution. *ACS Nano* **7**, 6156-6161 (2013).
26. S. Jiang *et al.*, Ultrahigh-performance transparent conductive films of carbon-welded isolated single-wall carbon nanotubes. *Sci. Adv.* **4**, eaap9264 (2018).
27. R. Saito, M. Fujita, G. Dresselhaus, M. S. Dresselhaus, Electronic-Structure of Chiral Graphene Tubules. *Appl. Phys. Lett.* **60**, 2204-2206 (1992).
28. G. G. Samsonidze *et al.*, Interband optical transitions in left- and right-handed single-wall carbon nanotubes. *Phys. Rev. B* **69**, 205402 (2004).
29. J. W. G. Wildoer, L. C. Venema, A. G. Rinzler, R. E. Smalley, C. Dekker, Electronic structure of atomically resolved carbon nanotubes. *Nature* **391**, 59-62 (1998).
30. G. Y. Ao, J. K. Streit, J. A. Fagan, M. Zheng, Differentiating Left- and Right-Handed Carbon Nanotubes by DNA. *J. Am. Chem. Soc.* **138**, 16677-16685 (2016).
31. Y. B. Chen *et al.*, Helicity-dependent single-walled carbon nanotube alignment on graphite for helical angle and handedness recognition. *Nat. Commun.* **4**, 2205 (2013).
32. G. Liu *et al.*, Simultaneous Discrimination of Diameter, Handedness, and Metallicity of Single-Walled Carbon Nanotubes with Chiral Diporphyrin Nanocalipers. *J. Am. Chem. Soc.* **135**, 4805-4814 (2013).
33. X. Peng *et al.*, Optically active single-walled carbon nanotubes. *Nat. Nanotechnol.* **2**, 361-365 (2007).
34. M. Yankowitz *et al.*, Emergence of superlattice Dirac points in graphene on hexagonal boron nitride. *Nat. Phys.* **8**, 382-386 (2012).
35. S. H. Zhao *et al.*, Observation of Drastic Electronic-Structure Change in a One-Dimensional

Moire Superlattice. *Phys. Rev. Lett.* **124**, 106101 (2020).
36. Z. Liu *et al.*, Determination of optical isomers for left-handed or right-handed chiral double-wall carbon nanotubes. *Phys. Rev. Lett.* **95**, 187406 (2005).
37. M. G. Burdanova *et al.*, Ultrafast Optoelectronic Processes in 1D Radial van der Waals Heterostructures: Carbon, Boron Nitride, and MoS2 Nanotubes with Coexisting Excitons and Highly Mobile Charges. *Nano Lett.* **20**, 3560-3567 (2020).
38. P. Y. K. Wang *et al.*, Enhanced In-Plane Thermal Conductance of Thin Films Composed of Coaxially Combined Single-Walled Carbon Nanotubes and Boron Nitride Nanotubes. *ACS Nano* **14**, 4298-4305 (2020).
39. C. Hu, V. Michaud-Rioux, W. Yao, H. Guo, Theoretical Design of Topological Heteronanotubes. *Nano Lett.* **19**, 4146-4150 (2019).
40. M. Liu *et al.*, Photoluminescence from Single-Walled MoS2 Nanotubes Coaxially Grown on Boron Nitride Nanotubes. *ACS Nano*, **15**, 8418-8426 (2021).
41. C. Liu *et al.*, One-Dimensional van der Waals Heterostructures as Efficient Metal-Free Oxygen Electrocatalysts. *ACS Nano* **15**, 3309-3319 (2021).
42. P. H. Ying, J. Zhang, Y. Du, Z. Zhong, Effects of coating layers on the thermal transport in carbon nanotubes-based van der Waals heterostructures. *Carbon* **176**, 446-457 (2021).
43. Y. Feng *et al.*, One-Dimensional van der Waals Heterojunction Diode. *ACS Nano* **15**, 5600-5609 (2021).
44. Y. Gogotsi, B. I. Yakobson, Nested hybrid nanotubes. *Science* **367**, 506-507 (2020).
45. S. Maruyama, R. Kojima, Y. Miyauchi, S. Chiashi, M. Kohno, Low-temperature synthesis of high-purity single-walled carbon nanotubes from alcohol. *Chem. Phys. Lett.* **360**, 229-234 (2002).
46. T. Inoue *et al.*, Effect of Gas Pressure on the Density of Horizontally Aligned Single-Walled Carbon Nanotubes Grown on Quartz Substrates. *J. Phys. Chem. C* **117**, 11804-11810 (2013).
47. A. Kinaci, J. B. Haskins, C. Sevik, T. Cagin, Thermal conductivity of BN-C nanostructures. *Phys. Rev. B* **86**, 115410 (2012).
48. T. Maaravi, I. Leven, I. Azuri, L. Kronik, O. Hod, Interlayer Potential for Homogeneous Graphene and Hexagonal Boron Nitride Systems: Reparametrization for Many-Body Dispersion Effects. *J. Phys. Chem. C* **121**, 22826-22835 (2017).
49. W. G. Ouyang, D. Mandelli, M. Urbakh, O. Hod, Nanoserpents: Graphene Nanoribbon Motion on Two-Dimensional Hexagonal Materials. *Nano Lett.* **18**, 6009-6016 (2018).
50. W. G. Ouyang *et al.*, Mechanical and Tribological Properties of Layered Materials under High Pressure: Assessing the Importance of Many-Body Dispersion Effects. *J. Chem. Theory Comput.* **16**, 666-676 (2020).
51. S. Plimpton, Fast Parallel Algorithms for Short-Range Molecular-Dynamics. *J. Comput. Phys.* **117**, 1-19 (1995).
52. http://lammps.sandia.gov.